\documentclass[useAMS,usenatbib]{mn2e}

\usepackage{graphicx}

\newcommand{\xmm}{{\it XMM-Newton}}

\newcommand{\kepler}{{\it Kepler}}

\title[Fast optical variability of V1504\,Cyg]{Differences in the fast optical variability of the dwarf nova V1504\,Cyg between quiescence and outbursts detected in \kepler\ data and simulations of the rms-flux relations.}
\author[A. Dobrotka]{
A. Dobrotka$^1$\thanks{E-mail: andrej.dobrotka@stuba.sk},
J.-U. Ness$^2$\\
$^1$Advanced Technologies Research Institute, Faculty of Materials Science and Technology in Trnava, Slovak University of Technology\\
in Bratislava, Paul\'inska 16, 91724 Trnava, Slovak Republic\\
$^2$XMM-Newton Science Operations Center, European Space Astronomy Center, PO Box 78, 28691 Villanueva de la Ca\~nada, Madrid,\\
Spain\\
}

\begin{document}

\date{Accepted ???. Received ???; in original form \today}

\pagerange{\pageref{firstpage}--\pageref{lastpage}} \pubyear{2015}

\maketitle

\label{firstpage}

\begin{abstract}
An optical light curve of SU\,UMa type dwarf nova V1504\,Cyg taken by \kepler\ was analysed in order to study fast optical variability (flickering). We calculated power density spectra and rms-flux relations for two different stages of activity, i.e. quiescence and regular outbursts. A multicomponent power density spectrum with two break frequencies was found during both activity stages. The rms-flux relation is obvious only in the quiescent data. However, while the collection of all outburst data do not show this variability, every individual outburst does show it in the majority of cases keeping the rms value approximately in the same interval. Furthermore, the same analysis was performed for light curve subsamples taken from the beginning, middle and the end of the supercycle both for quiescence and regular outbursts. Every light curve subsample shows the same multicomponent power density spectrum. The stability of the break frequencies over the supercycle can be confirmed for all frequencies except for the high break frequency during outburst, which shows variability, but with rather low confidence. Finally, the low break frequency can be associated with the geometrically thin disc or its inner edge, while the high break frequency can originate from the inner geometrically thick hot disc. Furthermore, with our statistical method to simulate flickering light curves, we show that the outburst flickering light curve of V1504\,Cyg needs an additional constant flux level to explain the observed rms-flux behaviour. Therefore, during the outbursts another non-turbulent radiation source should be present.
\end{abstract}


\section{Introduction}
\label{introduction}

Dwarf novae (DNe) are a subclass of interacting binaries with a white dwarf as a primary star and a main sequence star as a secondary. The latter is filling its Roche lobe and mass is transfered to the white dwarf. With or without weak magnetic fields of the primary, a geometrically thin optically thick accretion disc is formed. This class of interacting binaries is known as cataclysmic variables (CVs) (see e.g. \citealt{warner1995} for a review). The accretion disc is the main source of variability on large intervals of time scales and energies. DNe exhibit typical alternating quiescent and active (outburst) stages. The driving mechanism of this alternation is the viscous-thermal disc instability, generated by variations in ionization state of hydrogen (\citealt{osaki1974}, \citealt{hoshi1979}, \citealt{meyer1981}, \citealt{lasota2001}). During quiescence the mass accretion rate through the disc is much lower than during outbursts. While the mass is driven inwards, the angular momentum is transported outwards. Therefore, the different activity stages have different angular momentum redistribution and disc structure. Different typical time scales of variability patterns are expected.

An underlying accretion process is usually manifested in fast stochastic variability (sometimes also called flickering). Such oscillations are observed in optical and X-ray light curves of active galactic nuclei, X-ray binaries and CVs. This variability reflects as red noise in power density spectra (PDS). The shape of such PDS can be complicated by several characteristic break frequencies (see e.g. \citealt{sunyaev2000}, \citealt{scaringi2012b}, \citealt{dobrotka2014}). The best example up to date in CVs is detected in high-cadence optical \kepler\ data of the nova-like system MV\,Lyr, taken almost continuously over 600 days (\citealt{scaringi2012b}). The PDS show 4 different break frequencies.

Second and even more fundamental characteristic of flickering is the correlation between variability amplitude and flux over a wide-range of timescales. This rms-flux relation has a typical linear trend simply meaning that the higher the flux, the larger the degree of variability. Standard shot-noise models (e.g. \citealt{terrell1972}) with constantly decaying superposed flares (so-called stationary) seek to explain the broadband noise variability including the linear rms-flux relation. \citet{uttley2001} showed that the broadband variability of the black hole binary Cyg X-1 and the accreting millisecond pulsar SAX\,J1808.4-3658 are intrinsically non-stationary, and as a consequence the rms scales linearly with the flux. Therefore, this linearity is a fundamental characteristic of flickering activity. Following further studies of stochastic X-ray variability of X-ray binaries and active galaxies by \citet{uttley2005}, the linear rms-flux relation implies that short and long variability time scales are coupled multiplicatively, and the flux of stationary data should have a log-normal distribution.

This rms-flux linear relation is clearly a ubiquitous feature in X-ray binaries and active galaxies (\citealt{heil2012}). But to detect such variability in CVs, long enough observations are needed because of longer variability time scales. This is now possible with the \kepler\ mission. So far, the rms-flux relation in CVs was detected in three systems, MV\,Lyr (\citealt{scaringi2012a}), KIC\,8751494 and V1504\,Cyg (\citealt{vandesande2015}). The most promising physical model describing the rms-flux relation are variations in the accretion rate that are produced at different disc radii (\citealt{lyubarskii1997}, \citealt{kotov2001}, \citealt{arevalo2006}, \citealt{ingram2013}). This statement agrees with the recent modelling of the MV\,Lyr flickering by \citet{dobrotka2015}. The authors used their statistical method based on an unstable turbulent mass accretion rate (\citealt{dobrotka2010}), and found that the linear rms-flux relation with log-normally distributed flux of stationary data is an inherent characteristic of their method. 

V1504\,Cyg is a member of a specific subset of the DNe showing also considerably larger superoutbursts that occur in cycles, in addition to the regular outbursts (see e.g. \citealt{warner1995}, \citealt{lasota2001} for a review). The physical origins of these two types of outbursts are not the same. Two different physical scenarios have been proposed as explanations of the superoutburst, i.e. enhanced mass transfer or tidal thermal instability (\citealt{schreiber2004}). The former scenario suggests that the superoutbursts are generated when the disc mass exceeds a critical value, while the latter is based on the outer disc radius expanding to a certain critical radius (3:1 resonance radius) where the tidal activity is trigerring the superoutburst. \citet{osaki2013} analysed the same data as we use in this paper, and studied the appearance of superhumps (a variability connected to superoutbursts). The authors concluded that the superoutburst was initiated by a tidal instability (as evidenced by the growing superhump).

Finally, \citet{vandesande2015} studied the rms-flux relation of V1504\,Cyg during quiescence using the same data. The purpose of our work is to expand the analysis to the outburst stage and investigate the behaviour between two successive superoutbursts, i.e. during the supercycle. After the presentation of the analysed observations in Section~\ref{data} we present studies of the power density spectra (Section~\ref{pds_analysis}) and the rms-flux relation (Section~\ref{rms_analysis}). The results are discussed and summarized in Sections~\ref{discussion} and \ref{summary} respectively.

\section{Data}
\label{data}

The NASA \kepler\ mission offers a unique opportunity to study fast optical variability of a variety of accreting objects. The spacecraft was continuously monitoring a 116 square degree field-of-view with a cadence of 58.8\,s until a second reaction wheel failure occurred. Such an almost continuous\footnote{The satellite performed quarterly $90\deg$ rolls in order to have the solar panels face to the Sun.} observation of the same sky region yields high quality optical light curves of several CVs over a time span of hundreds of days\footnote{For \kepler\ CVs light curves summary see \citet{vandesande2015}}. This allows detailed studies of the power density spectra (PDS) over a wide frequency range.

For our detailed light curve analysis we selected one of the longest light curves of a CV, i.e. the SU\,UMa DN system V1504\,Cyg (KIC number 7446357, orbital period of 1.67\,h). The light curve duration of about 1400 days consists of quiescent emission with 118 fully covered regular outbursts distributed into 11 full supercycles. The fast stochastic variability and the orbital pattern are obvious in both quiescence and outburst.

\begin{figure*}
\includegraphics[width=67mm,angle=-90]{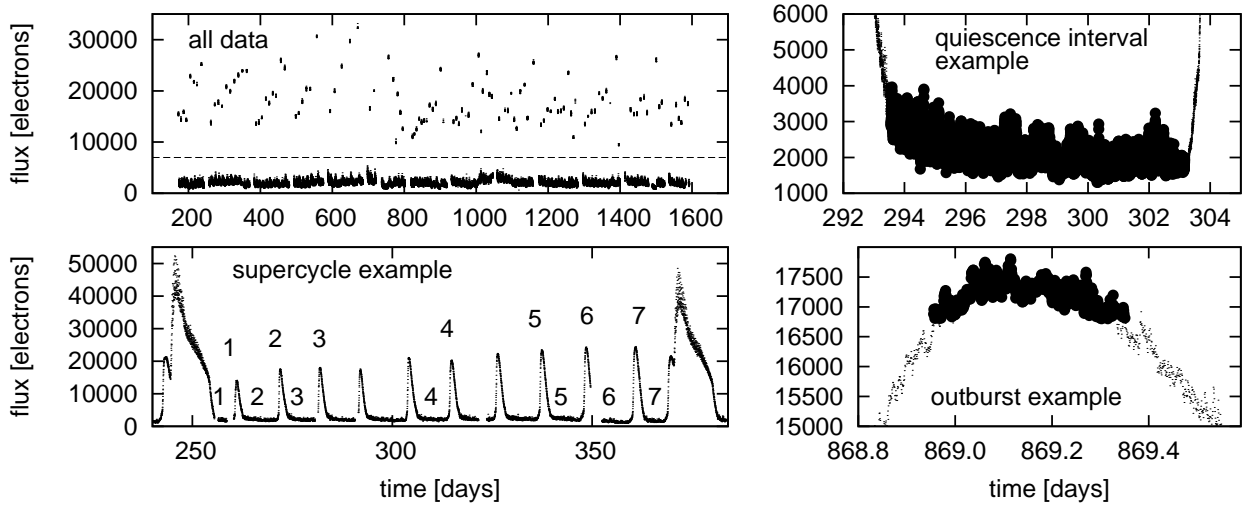}
\caption{Analysed light curves. Upper left panel: all data divided into quiescent (below the dashed line) and outburst subsamples (above the dashed line). Lower left panel: supercycle example with marked supercycle subsamples, i.e. 1 - start, 2 - start+1, 3 - start+2, 4 - middle, 5 - end-2, 6 - end-1 and 7 - end (see text for details). Right panels: examples of quiescent and outburst intervals. The dots show the original light curve and circles are the selected data (outburst or quiescence).}
\label{lcs}
\end{figure*}

We first extracted separate light curves from periods of quiescence and outburst. We first identified all outburst peaks with a simple algorithm. Subsequently, we determined the times when the rise to outbursts ended and when the decline from the outburst started by taking the time derivatives of flux and applying certain limits to the derivatives after and before the peak (derivative limit = 1000 electrons/s). Outburst light curves were extracted from the times in between these time intervals. Time intervals of quiescence between outbursts were identified by applying an upper flux limit (dashed curve in Fig.~\ref{lcs}) in between the times of decline of one outburst and rise into the respective next outburst again determined by a derivative limit (derivative limit = 1500 electrons/s). The limits on the derivatives and flux were chosen visually. Both light curves with details are shown in Fig.~\ref{lcs} and all starting and ending times of all individual light curve subsamples are summarized in Table~\ref{interval_times}.

Subsequently we selected the start, middle and end stage of the supercycle in order to study any evolution for both quiescence and outbursts. Therefore, we collected all quiescent intervals lasting as first (later called start), second (start+1) and third (start+2) after the superoutburst. A similar selection was done with quiescent intervals before the superoutburst, i.e. we collected last (end), second (end-1) and third (end-2) interval before the superoutburst. As middle of the supercycle we choose the closest interval to the half time of the supercycle. The same was performed with the outburst intervals (inset of Fig.~\ref{lcs}).
\begin{table*}
\caption{Starting and ending dates of individual light curve intervals for all selected quiescence and outburst subsamples. The number of full supercycles is 11 which yields 11 middle light curve intervals, while part of the supercycle at the beginning and at the end of the light curve yields 12 light curve intervals in the case of other subsamples. Where an observing gap is interrupting the light curve, or the searched interval is somehow not distinguishable, the light curve interval is missing.}
\begin{center}
\begin{tabular}{lcccclcccc}
\hline
\hline
supercycle & quiesc. & quiesc. & outb. & outb. & supercycle & quiesc. & quiesc. & outb. & outb.\\
position & start & end & start & end & position & start & end & start & end\\
& (days) & (days) & (days) & (days) & & (days) & (days) & (days) & (days)\\
\hline
start & 256.3280 & 258.4775 & 260.8069 & 261.1876 & end-2 & 215.8605 & 220.9621 & 213.5107 & 213.8853\\
& 381.4602 & 385.6599 & 386.3908 & 386.7041 & & 339.7604 & 346.9603 & 337.1763 & 337.557\\
& 491.8608 & 496.3605 & 496.9292 & 497.2998 & & 444.4605 & 454.3602 & 442.0567 & 442.2127\\
& 586.3605 & 591.7604 & 592.2522 & 592.6248 & & -- & -- & -- & --\\
& 696.4608 & 702.1602 & 702.7916 & 703.1669 & & -- & -- & -- & --\\
& 817.6604 & 820.3605 & 821.3372 & 821.7221 & & 782.5609 & 786.1599 & 779.9269 & 780.3152\\
& 930.1603 & 934.6599 & 935.2852 & 935.6638 & & 888.4603 & 898.3601 & 886.2139 & 886.7254\\
& 1057.0605 & 1061.5598 & 1062.2777 & 1062.6557 & & 1030.3605 & 1033.3600 & 1027.9310 & 1028.2879\\
& 1173.1603 & 1177.9617 & 1178.4786 & 1178.8008 & & -- & -- & 1133.0365 & 1133.4343\\
& 1295.6785 & 1300.0602 & 1300.6895 & 1300.9974 & & 1264.3604 & 1267.9602 & 1262.0771 & 1262.8127\\
& -- & -- & -- & -- & & -- & -- & 1371.3064 & 1371.3323\\
& 1534.9608 & 1540.0604 & 1540.7157 & 1541.2333 & & 1501.9603 & 1506.1602 & 1499.0002 & 1499.3572\\
& & & & & & & & & \\
start+1 & 262.6602 & 271.0581 & 271.7412 & 271.9762 & end-1 & 224.5604 & 233.2603 & 222.3209 & 222.6009\\
& 388.3606 & 393.4601 & 394.1099 & 394.5485 & & 352.4397 & 359.5599 & 348.2367 & 348.6249\\
& 499.0601 & 504.6044 & 505.0179 & 505.3700 & & 458.2604 & 464.5603 & 455.2695 & 455.7606\\
& 594.1606 & 604.0599 & 604.8896 & 605.2485 & & 539.5853 & 553.9600 & 536.9200 & 537.2878\\
& 704.8580 & 713.8601 & 714.7352 & 714.9627 & & 649.9608 & 667.9600 & 647.3603 & 647.7540\\
& 822.7608 & 825.7599 & 826.5732 & 826.8824 & & 790.0607 & 793.3602 & 787.5201 & 788.1236\\
& 936.9335 & 942.4601 & 943.2925 & 943.6842 & & 902.2603 & 905.9364 & 899.4737 & 899.9335\\
& 1064.2610 & 1070.0155 & -- & -- & & 1037.2606 & 1040.5604 & 1035.0144 & 1035.6015\\
& 1180.3606 & 1185.7600 & 1186.5957 & 1186.9240 & & 1143.1606 & 1147.0601 & 1140.9021 & 1141.6622\\
& 1302.4604 & 1308.7603 & 1309.5906 & 1309.8256 & & 1270.3605 & 1272.4604 & 1268.8763 & 1269.1440\\
& 1419.1587 & 1426.9598 & 1427.7765 & 1428.1694 & & 1390.9608 & 1394.2600 & 1388.1697 & 1388.5763\\
& 1542.7584 & 1553.5598 & 1554.1177 & 1554.5019 & & 1509.7605 & 1513.3602 & 1507.2479 & 1507.7710\\
 & & & & & & & & &\\
start+2 & 273.7607 & 280.5307 & 281.6477 & 281.9808 & end & 236.8606 & 241.6584 & 234.4168 & 234.8459\\
& -- & -- & 402.5966 & 402.9849 & & 363.1602 & 368.2603 & 360.5455 & 360.9752\\
& 506.8611 & 514.3598 & 514.9804 & 515.3236 & & 468.4606 & 478.6601 & 466.1433 & 466.5820\\
& 607.0575 & 623.5601 & -- & -- & & 558.1605 & 573.4600 & 555.1349 & 555.4414\\
& 716.8602 & 719.5580 & -- & -- & & 672.1602 & 683.5603 & 669.0184 & 669.3691\\
& 828.1603 & 831.1600 & 832.2969 & 832.7492 & & 796.3605 & 800.26001 & 794.0964 & 794.4377\\
& 945.1608 & 949.3600 & 950.2032 & 950.6010 & & 909.4606 & 916.9599 & 907.2884 & 907.4853\\
& 1072.3456 & 1077.7603 & 1078.4399 & 1078.8309 & & 1044.4604 & 1045.0604 & 1041.7701 & 1042.5397\\
& 1188.4607 & 1195.9601 & 1196.7468 & 1197.1276 & & 1148.8602 & 1152.7556 & 1157.7912 & 1158.3014\\
& 1311.4603 & 1318.3601 & 1319.0753 & 1319.4240 & & 1276.0603 & 1281.4603 & 1274.1299 & 1274.2362\\
& 1429.9607 & 1436.3187 & 1437.0366 & 1437.4255 & & 1396.3605 & 1398.1586 & 1394.8117 & 1395.0637\\
& 1556.2606 & 1560.7616 & 1561.5885 & 1561.9712 & & 1516.9606 & 1522.0602 & 1514.3703 & 1515.0671\\
 & & & & & & & & &\\
middle & 306.4605 & 313.6603 & 314.5675 & 314.9619 & middle & 979.3605 & 987.7599 & 988.5288 & 988.9423\\
& 424.9608 & 431.8600 & 432.8933 & 433.2836 & & 1105.9603 & 1110.1566 & 1110.9556 & 1111.6421\\
& 528.7606 & 535.9603 & 526.3112 & 526.6729 & & 1219.3602 & 1224.4600 & 1225.6397 & 1226.0511\\
& 627.1603 & 646.3605 & 624.4659 & 624.8167 & & 1342.6604 & 1349.2604 & 1350.0273 & 1350.4074\\
& 736.0604 & 753.7604 & 754.5893 & 754.9625 & & 1467.1606 & 1472.5604 & 1465.0343 & 1465.4021\\
& 862.6604 & 868.0599 & 861.0216 & 861.3267 & & & & &\\
\hline
\end{tabular}
\end{center}
\label{interval_times}
\end{table*}

\section{PDS analysis}
\label{pds_analysis}

For the PDS analysis we applied the Lomb-Scargle algorithm (\citealt{scargle1982}) which can handle gaps in the light curves. The low frequency end of the studied PDSs usually depends on the light curve duration. In order to study differences between outburst and quiescence both PDSs must be comparable on the studied frequency interval. Therefore, we set this limit empirically (see next Section). The high frequency end is usually limited by the white noise or power rising of the PDS to the Nyquist frequency. We empirically defined this limit to 10$^{-2.2}$\,Hz. The quiescent data were divided into one day intervals. Every quiescent interval yields an individual PDS. The outburst data were not divided into shorter subsamples. Every outburst light curve was used for individual PDS calculation. Finally, a mean PDS was calculated from all individual PDSs. We subsequently binned the mean PDS into equally spaced bins and the mean values with the errors of the mean was fitted with a broken power law.

\subsection{All data}

\begin{figure*}
\includegraphics[width=91mm,angle=-90]{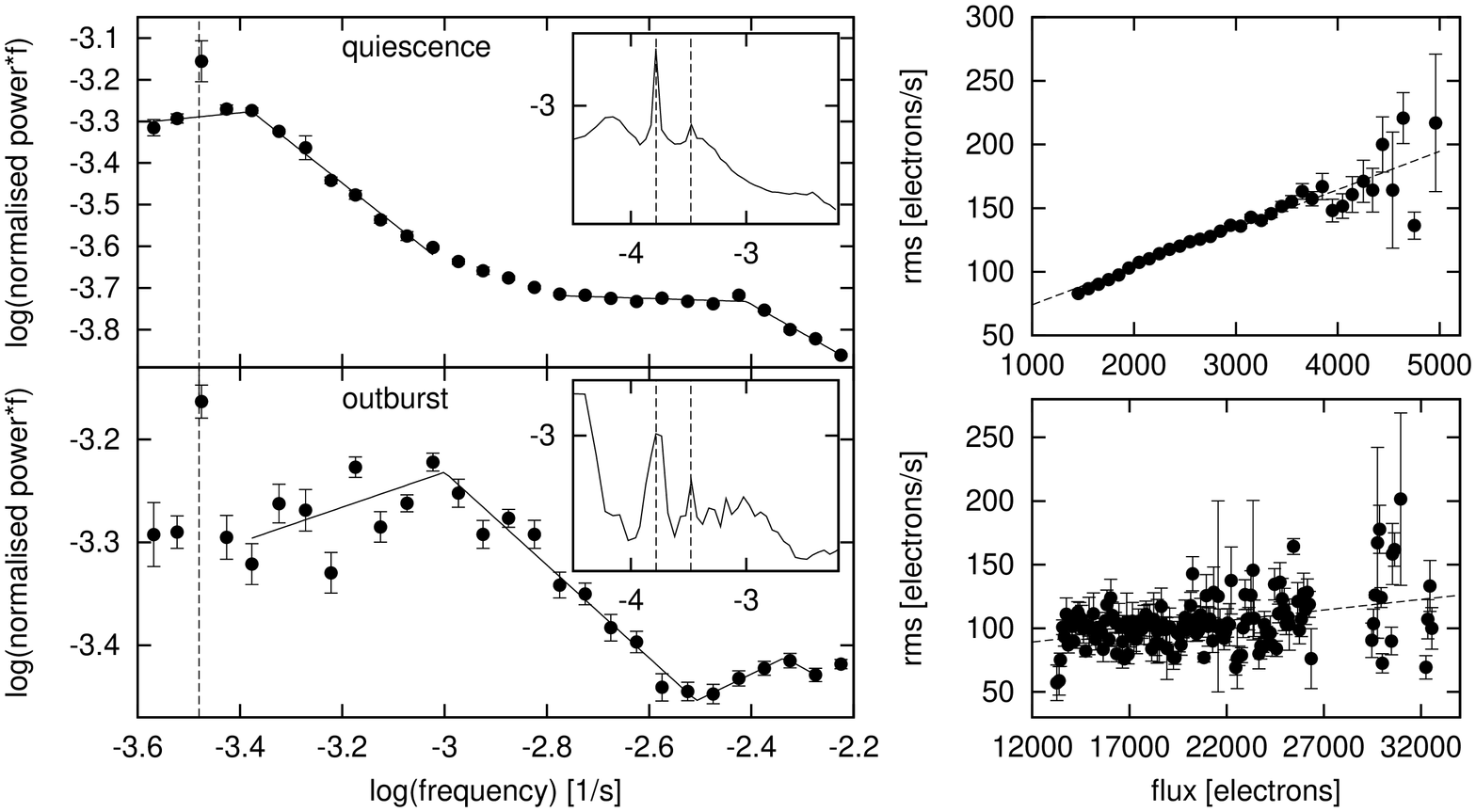}
\caption{Left column: mean PDS of quiescent and outburst data. Solid lines are the broken power law fits. The dashed lines shows the orbital frequency $f = 10^{-3.78}$\,Hz (1.67\,h) with its first harmonic $f = 10^{-3.48}$\,Hz (0.84\,h). The insets show low resolution mean PDSs with larger x-axis range. Right column: Rms-flux relations calculated from both quiescent and outburst data. The dashed lines are linear fits.}
\label{pds_all}
\end{figure*}

Fig.~\ref{pds_all} shows (left column) the PDSs of all quiescent and outburst data. The orbital frequency and its first harmonic can clearly be seen and are marked by dashed lines. While the quiescent PDS shows a significant peak around $f = 10^{-4.2}$\,Hz (see the low resolution insets in Fig.~\ref{pds_all}), the outburst PDS is influenced by the overall inherent outburst shape in this low frequency region and shows a steep rise toward lowest frequencies. A comparable behaviour of both PDSs starts above the orbital frequency. Therefore, in order to focus our study on the common behaviour of both PDSs, we choose the frequency $f = 10^{-3.6}$\,Hz as a low frequency end of the studied PDSs.

The multicomponent shape of the PDS with two break frequencies $f_{\rm L}$ (low frequency) and $f_{\rm H}$ (high frequency) is clearest in the quiescence PDS. The outburst PDS has one clear break frequency while a multicomponent shape with a second break frequency around $f = 10^{-2.3}$\,Hz may be possible. We fitted the binned PDSs with broken power law models\footnote{The bin containing the orbital harmonic frequency was excluded from the fitting procedure.}. In the case of outburst we used a multicomponent model\footnote{We reduced also the fitting frequency interval because of large scatter in the data in the lowest frequency region which influenced the iteration procedure in some cases. We used $f = 10^{-3.4}$\,Hz as the low-frequency end instead of $f = 10^{-3.6}$\,Hz. This does not influence the study, because the omitted low-frequency region is useless and the searched break frequency is clearly around $f = 10^{-3.0}$\,Hz.}, while in the quiescent PDS two separate broken power law fits were more adequate due to the convex shape between the two broken power laws. The resulting fitted PDS parameters with 1-$\sigma$ errors (derived from $\chi^2$ test) are summarized in Table~\ref{fitted_frekv}. In the multicomponent fit of the outburst PDS, we had problems with the fitting procedure while keeping $f_{\rm H}$ constant in order to get the $\chi^2$ curve. The fit often converged to deformed unacceptable solutions. Therefore, in order to derive the 1-$\sigma$ error of the outburst $f_{\rm H}$, we fitted a single power law model to the selected PDS interval. Finally, the $\chi^2$ residuals of the fits often yield asymmetric curves. Therefore, in Table~\ref{fitted_frekv} we show lower and upper limits for clarity, instead of plus/minus errors.

In general, we can conclude that the multicomponent shape of the PDS with two characteristic break frequencies is present in both activity stages. Furthermore, $f_{\rm L}$ is significantly different, while $f_{\rm H}$ agree within the 1-$\sigma$ errors for both activity stages.
\begin{table*}
\caption{Fitted low and high ($f_{\rm L}$ and $f_{\rm H}$) break frequencies and $a$ coefficient from the rms-flux linear fit for different subsamples with 1-$\sigma$ limits. "Stage" is the activity stage following the disc instability model and "sample" is the analysed data subsample. Two solutions are listed for the subsample "outburst, end" because it is model dependent. The outburst 1-$\sigma$ limits within parentheses are derived from single broken power law fits over a reduced frequency interval, instead of the original multicomponent fit performed over the full frequency interval.}
\begin{center}
\begin{tabular}{llccccccl}
\hline
\hline
stage & sample & & log($f_{\rm L}$) & & & log($f_{\rm H}$) & & $a$\\
\hline
quiescence & all & -3.37 & -3.38 & -3.39 & -2.39 & -2.41 & -2.42 & 0.030 $\pm$ 0.002\\
outburst & all & -2.94 & -3.00 & -3.03 & (-2.27) & -2.33 & (-2.40) & 0.0017 $\pm$ 0.0003\\
\\
quiescence & start & -3.36 & -3.39 & -3.42 & -2.35 & -2.39 & -2.45 & 0.030 $\pm$ 0.003\\
quiescence & start+1 & -3.36 & -3.41 & -3.46 & -2.38 & -2.40 & -2.43 & 0.029 $\pm$ 0.003\\
quiescence & start+2 & -3.34 & -3.42 & -3.46 & -2.40 & -2.42 & -2.44 & 0.046 $\pm$ 0.004\\
quiescence & middle & -3.26 & -3.30 & -3.42 & -2.34 & -2.39 & -2.43 & 0.056 $\pm$ 0.003\\
quiescence & end-2 & -3.32 & -3.37 & -3.40 & -2.35 & -2.39 & -2.43 & 0.029 $\pm$ 0.004\\
quiescence & end-1 & -3.38 & -3.40 & -3.45 & -2.33 & -2.42 & -2.45 & 0.044 $\pm$ 0.003\\
quiescence & end & -3.33 & -3.43 & -3.47 & -2.37 & -2.40 & -2.44 & 0.050 $\pm$ 0.008\\
\\
outburst & start & -2.97 & -3.07 & -3.18 & (-2.39) & -2.43 & (-2.44) & 0.001 $\pm$ 0.002\\
outburst & start+1 & -2.83 & -2.94 & -3.16 & -- & -2.33 & -- & 0.002 $\pm$ 0.001\\
outburst & start+2 & -2.79 & -2.92 & -3.03 & (-2.44) & -2.47 & (-2.51) & 0.003 $\pm$ 0.001\\
outburst & middle & -2.72 & -2.90 & -3.10 & -- & -2.52 & -- & 0.0018 $\pm$ 0.0006\\
outburst & end-2 & -2.73 & -2.95 & -3.10& (-2.28) & -2.32 & (-2.34) & 0.0017 $\pm$ 0.0005\\
outburst & end-1 & -2.72 & -2.82 & -3.14 & (-2.33) & -2.41 & (-2.44) & 0.0022 $\pm$ 0.0004\\
outburst & end & -- & -2.59 & -- & -- & -2.29 & (-2.46) & 0.0018 $\pm$ 0.0005\\
 & & -- & -3.01 & -- & & & &\\
\hline
\end{tabular}
\end{center}
\label{fitted_frekv}
\end{table*}

\subsection{Supercycle subsamples}

\begin{figure*}
\includegraphics[width=160mm,angle=-90]{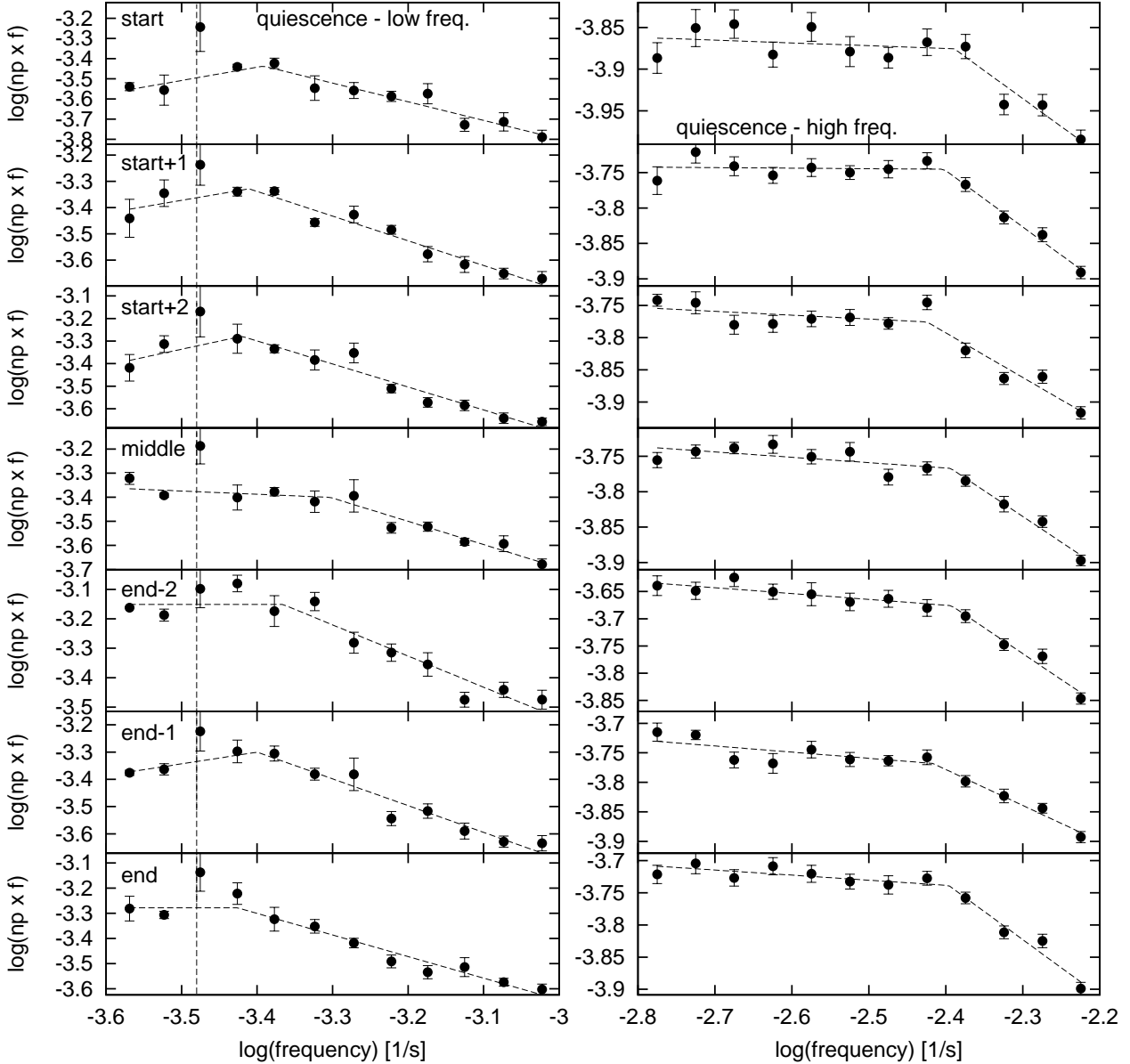}
\caption{Power density spectra of quiescent data with broken power law fits for different supercycle stages. The vertical dashed line shows the first harmonic of the orbital frequency.}
\label{pds_quiesc}
\end{figure*}

\begin{figure*}
\includegraphics[width=160mm,angle=-90]{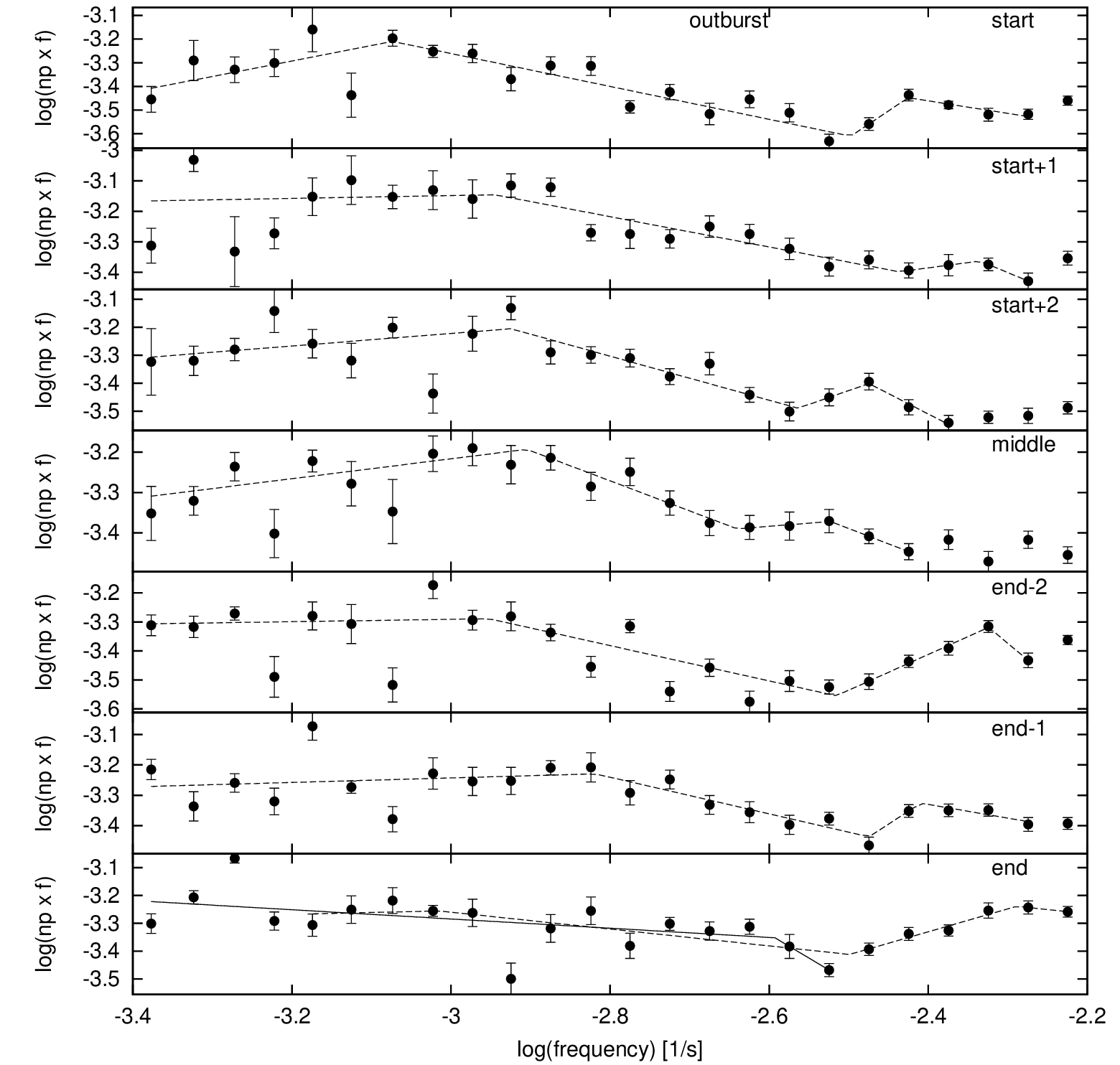}
\caption{Same as Fig.~\ref{pds_quiesc} derived from the outburst light curves.}
\label{pds_outb}
\end{figure*}

We performed the same analysis for other data subsamples collected for different supercycle phases (11 full covered supercycles). Fig.~\ref{pds_quiesc} shows low and high frequency parts with the broken power law fits separately for quiescence, and Fig.~\ref{pds_outb} shows the outburst PDSs with the multicomponent fits. Clearly $f_{\rm L}$ and $f_{\rm H}$ are present in all supercycle phases in both cases. The outburst PDSs show considerably larger scatter than the quiescence data, likely because of much less data. This is visible mainly in the last (end) PDS where it was difficult to find a best fit. The result is not robust against some factors, i.e. initial parameter estimate or the frequency interval over which the fit was performed. We show two solutions, i.e. the multicomponent fit as it converged, and visually motivated simple broken power law. Furthermore, in the case of outburst $f_{\rm H}$ the $\chi^2$ residuals calculations did not yield the searched 1-$\sigma$ confidence limits in every case. Considerable reductions of the confidence limit are needed to determine the error from the $\chi^2$ residuals curves but such low confidence intervals are meaningless. Therefore, some fitted $f_{\rm H}$ values are questionable, and we can only conclude the presence of $f_{\rm H}$ but with a highly uncertain value. All fitted break frequencies with derived 1-$\sigma$ limits are again summarized in Table~\ref{fitted_frekv}.

\begin{figure*}
\includegraphics[width=160mm,angle=-90]{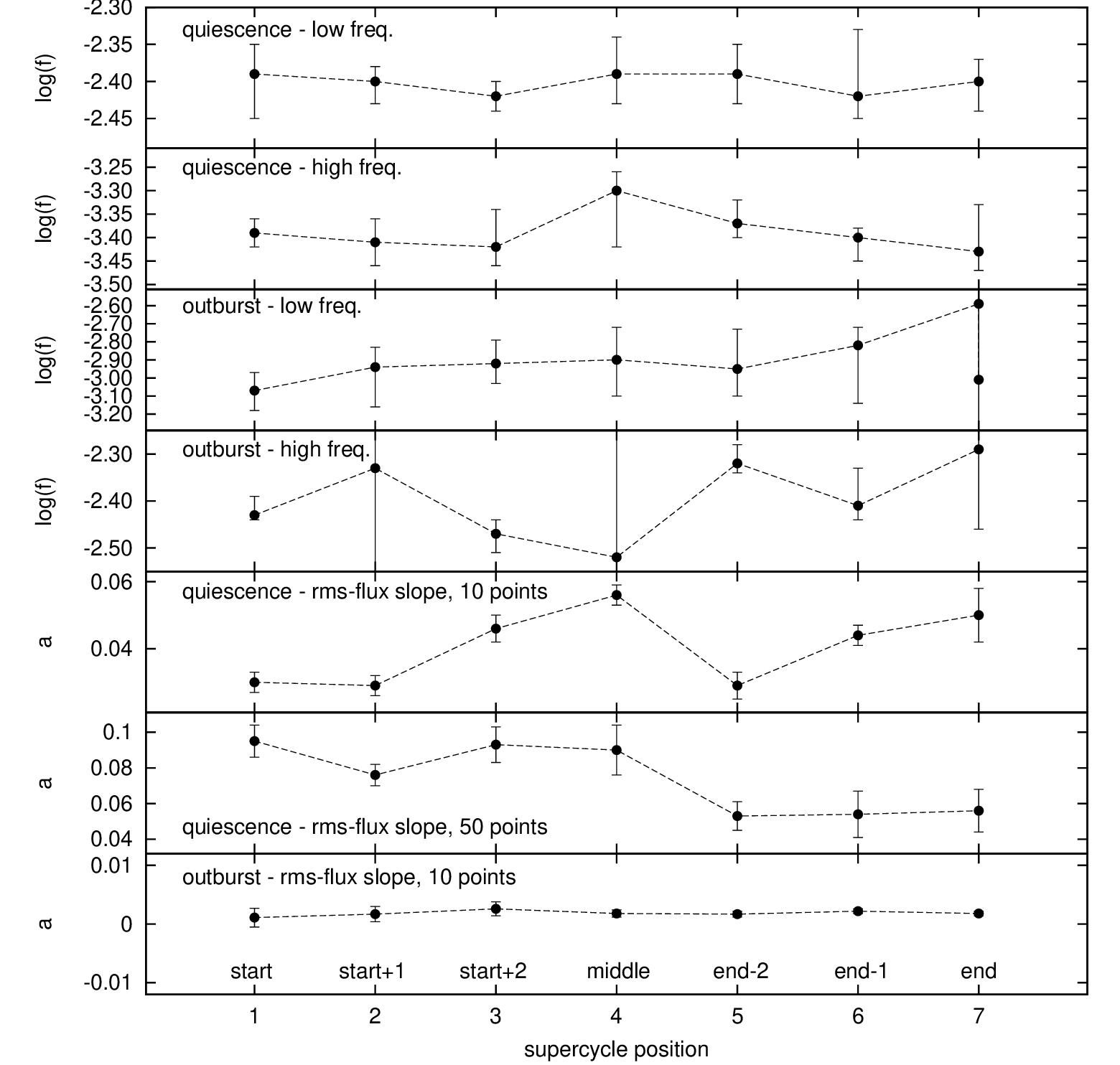}
\caption{Evolution of the fitted PDSs and rms-flux parameters during the supercycle. The error bars are 1-$\sigma$ errors. When this confidence limit was not reached in the $\chi^2$ test (confidence interval lower than the selected 1-$\sigma$), a vertical line over the whole y-axis interval is used instead.}
\label{evol}
\end{figure*}

Fig.~\ref{evol} shows the evolution of the fitted break frequencies as a function of the supercycle phase. All frequencies except the outburst $f_{\rm H}$ are stable within the derived 1-$\sigma$ confidence limits, and no trend can be deduced. Just in the case of outburst $f_{\rm L}$ a rising trend is possible but due to large errors still speculative. The outburst $f_{\rm H}$ values are scattered, and some values do not agree within the 1-$\sigma$ confidence interval. This suggests non-stability during the supercycle. But slightly larger confidence limits than the used 1-$\sigma$ would yield a match within the confidence frequency limits. Therefore, the instability of the outburst $f_{\rm H}$ is possible with rather low probability of approximately 1-$\sigma$.

\subsection{X-ray data}
\label{xray_try}

V1504\,Cyg was in the field of an \xmm\ observation taken 2014 October 15 00:45-07:08UT (ObsID 0743460201) that, at the time of writing is still proprietary. With the kind permission of the PI, J.H.M.M. Schmitt, we extracted the X-ray light curve of V1504\,Cyg from the two MOS detectors (the source was not within the field of view of the pn) in 50\,s binning and studied it in the same way as in \citet{dobrotka2014} but only found white noise. This may be due to the count rate of $<0.1$ counts per second in each MOS light curve, 100 times lower than in RU\,Peg which we tested by simulations. We took a subsample of the \kepler\ data and reduced the count rate to the same level as observed with \xmm. Subsequently we added Poisson noise to the simulated light curve in order to get a comparable flux histogram than the X-ray data. The break frequencies are still in the simulated light curve, but can not be recovered, supporting our interpretation that the count rate is too low. Thus, if the break frequencies are also present in the X-ray light curve, they would be buried in the noise.

While re-binning can increase the signal to noise ratio, it did not change the negative results. Furthermore, the larger bin size reduces the time resolution of the light curve and any binning larger han 100\,s is already larger than the required time resolution for a detection of $f_{\rm H}$. We thus conclude that V1504 Cyg is too faint in X-rays to complement our optical study with an X-ray PDS with the available sensitivity of the current instrumentation. 

\section{Rms-flux relation analysis}
\label{rms_analysis}

\subsection{Data}

The absolute rms amplitude of variability is defined as square-root of the variance, i.e.
\begin{equation}
\sigma_{\rm rms} = \sqrt{\frac{1}{N - 1} \sum^N_{i = 1} (x_i - \overline{x})^2},
\end{equation}
where $N$ is the number of data points, $x_i$ is the $i$-th point and $\overline{x}$ is the mean value of all $x_i$. We subdivided all light curves into small parts each containing 10 points (we tried more values, i.e. 25, 50, 100 and 200) of the light curve. Subsequently for each small part we calculated the corresponding rms and mean flux. The binned data by flux were fitted with a linear function.

The right column of the Fig.~\ref{pds_all} shows the rms-flux relations of all quiescent and outburst data. The quiescent rms-flux can be clearly seen as expected from \citet{vandesande2015}, but the outburst case does not show the same characteristics. The linear fit shows a rising trend, but over a wide flux range. The parameter $a$ of the linear function $rms = a \psi + b$ ($a$ is the gradient, $b$ is the vertical offset and $\psi$ is the flux) is only 0.002 compared to 0.03 in quiescence. Furthermore, the rms-flux data are always more scattered toward the highest flux, which can generate an artificial linear rising trend. We selected the less scattered data region with flux between 15000 and 20000 electrons/s, which yields $a = -0.001$. Clearly it is hard to talk about a rising linear trend which should be robust against flux range selection as is obvious in the quiescent case.

\begin{figure*}
\includegraphics[width=160mm,angle=-90]{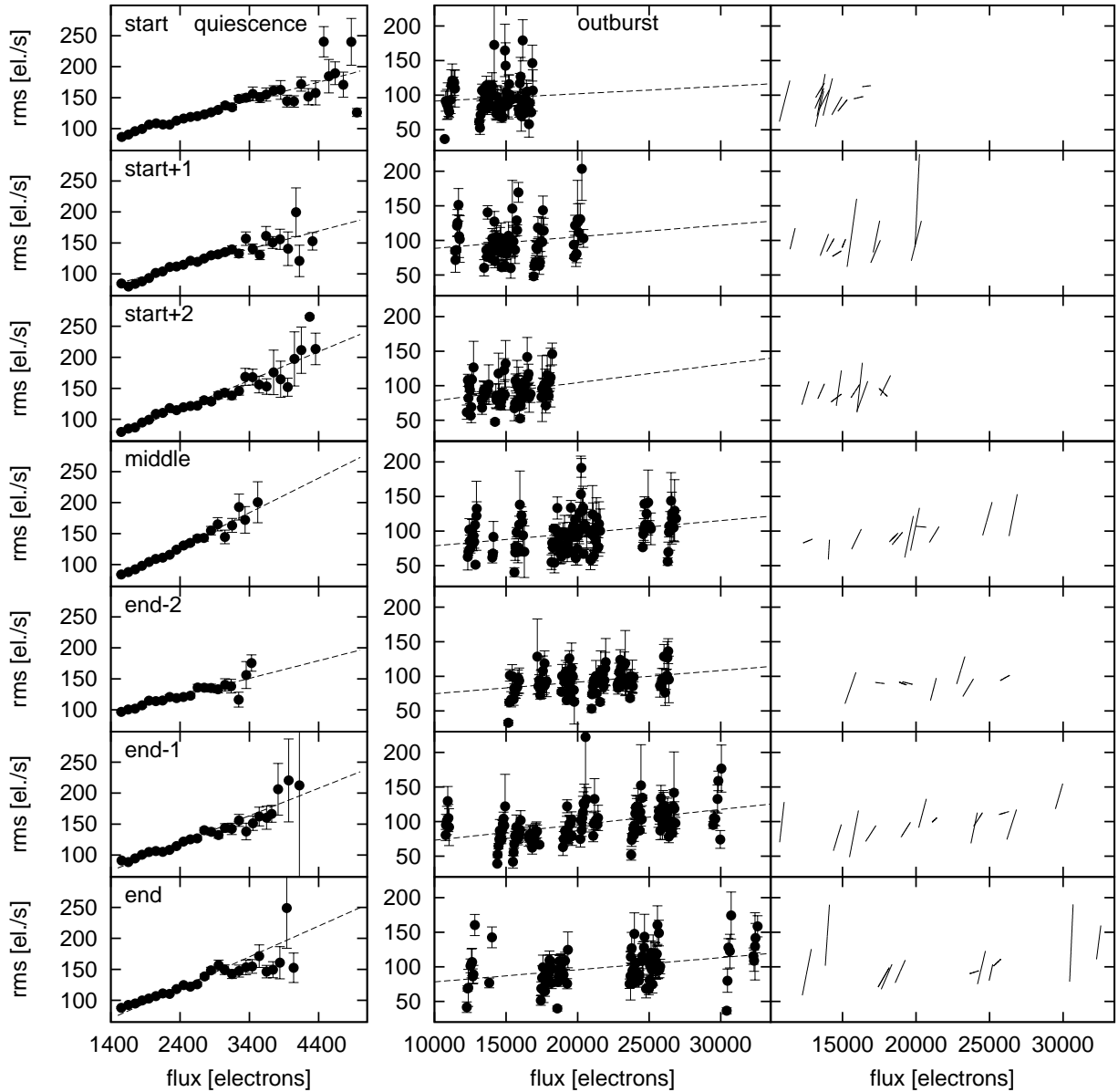}
\caption{Rms-flux relations with corresponding linear fits for quiescent and outburst data. The right column shows linear fits to individual outbursts.}
\label{rms_quiesc_outb}
\end{figure*}

We performed the same procedure for all other data subsamples and the resulting rms-flux relations are shown in Fig.~\ref{rms_quiesc_outb}. The basic difference between quiescent and outburst rms-flux is robust against the supercycle phase. The evolution of $a$ is shown in Fig.~\ref{evol} (rms calculated from 10 and 50 points is shown) and all $a$ values are listed in the last column of table~\ref{fitted_frekv}. The quiescent $a$ values are scattered in the supercycle without any significant trend, and the "evolution" shape is not robust against the selected number of data points over which the rms is calculated. All outburst $a$ values are still of the order of 0.001 (0.003 as maximum). Therefore, as for all outburst data, any linear rms-flux trend is not convincing.

In the last column of the Fig.~\ref{rms_quiesc_outb} we show linear fits to the rms-flux data of individual outbursts. Clearly, while the typical rms-flux relation is hardly detectable in the collection of all outburst data, the individual outbursts do show this relation in most cases. Moreover, linear fits of all individual outbursts lie approximately between 50 and 150 electrons/s (the rms-flux values are concentrated between 50 and 200 electrons/s).

\subsection{Simulations}

In \citet{dobrotka2010} we developed a statistical method to simulate light curves dominated by red noise, also known as flickering. The method is based on the simple idea of discrete angular momentum transport between two adjacent concentric rings in a geometrically thin, optically thick, hot ionized steady-state disc. The angular momentum is carried by discrete turbulent bodies with an exponential distribution of the dimension scales. When a turbulent body penetrates from one ring to the adjacent one, it changes the tangential velocity and distance from the center, while the mass is conserved. This changes the angular momentum. The sum of changes in angular momentum of all bodies must be equal to the total difference in angular momentum between two concentric rings. The distribution function of dimension scales is equal to the distribution function of turbulent event time scales because the dimension scale divided by the local viscous velocity gives the time scale of a turbulent event. All events liberate energy in the form of a flickering flare. The distribution functions between pairs of rings are summed over the entire disc (from the inner disc radius to the outer disc radius), which yields the final distribution function of flare time scales. Following this distribution function, flickering flares are randomly redistributed into a synthetic light curve of the same duration and sampling as the observations. Such synthetic light curves are subsequently analysed.

Furthermore, in \citealt{dobrotka2015} we applied the simulation method to the flickering study of MV\,Lyr system observed by \kepler. We showed that the simulated light curves show the typical linear rms-flux relation. MV\,Lyr is a nova-like system with the disc in the so-called hot stage with fully ionized hydrogen (see \citet{lasota2001} for a review). This hot stage, together with the steady state, is the essential assumption of our model. The studied data of V1504\,Cyg in this paper do not satisfy these basic assumptions, i.e. the quiescence is a cold stage with no ionized hydrogen, and the outburst is not in a steady state. Therefore, we do not use these simulations to study the PDS in detail, but we use it to study how it influences the rms-flux relation. The latter is based on flickering flare redistribution and the principle is robust against the underlying physical model of the accretion, i.e. non steady/steady state, quiescence/outburst etc...

\begin{figure*}
\includegraphics[width=110mm,angle=-90]{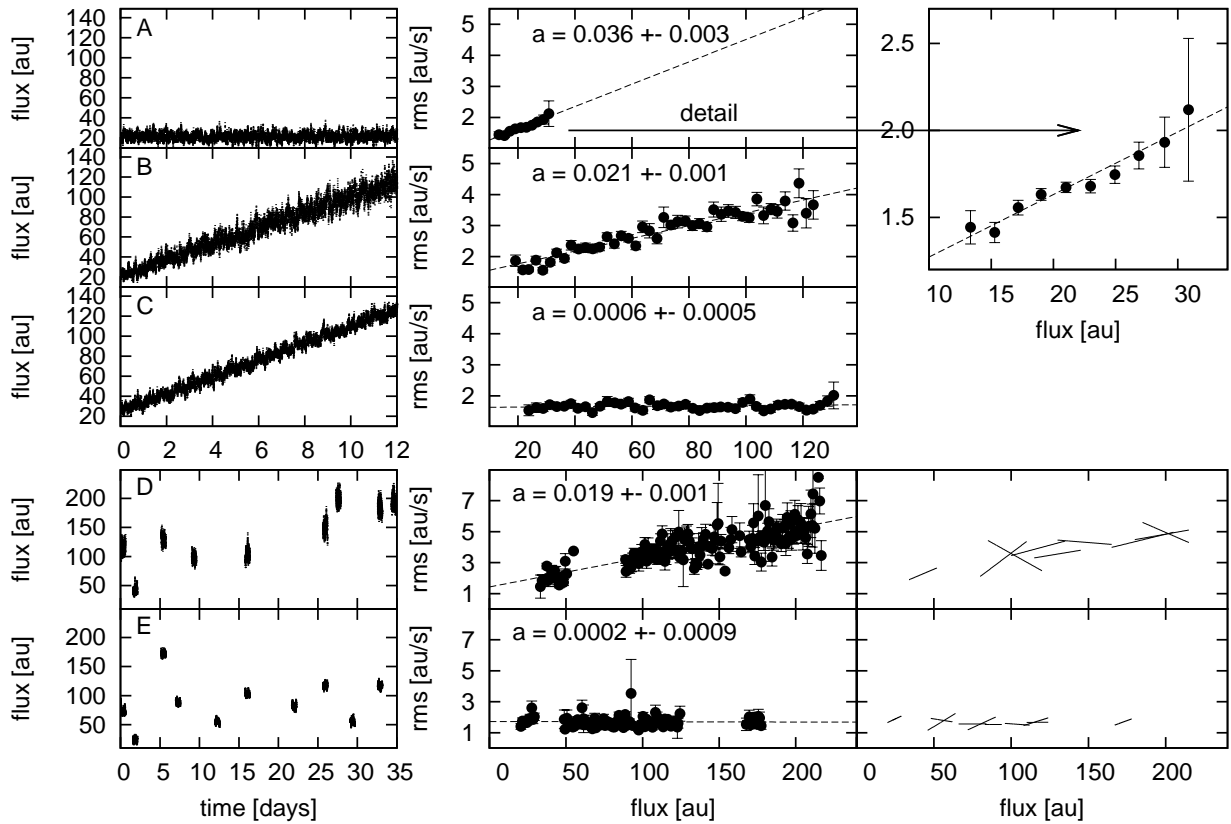}
\caption{Simulated light curves (left column) and corresponding rms-flux binned data. Middle column shows the rms-flux data taken from full light curves with linear fits as dashed lines with $a$ being linear coefficients of equation $rms = a \psi + b$ (where $\psi$ is the flux). The two lower right panels show linear fits to individual simulated light curve subsamples (outbursts). The upper right square panel is a detail of the upper middle panel. Flux is in arbitrary units (au). A - constant mass accretion rate, B - increasing mass accretion rate with correspondingly increasing number of flickering flares per light curve subsegment, C - the same as B but with constant flickering flare number per light curve subsegment and an additional vertical offset is added instead, D and E - the same as B and C respectively, but light curve subsegments are randomly redistributed in order to simulate the outbursts. See text for details.}
\label{rms_simul}
\end{figure*}

It appears that using a larger number of flares per light curve increases the rms and also the flux (\citealt{dobrotka2015}) while the PDS remains unchanged (\citealt{dobrotka2014}). In order to investigate how this plays a role in different rms-flux behaviour in quiescence and outburst of V1504\,Cyg, we simulated different light curves with red noise (Fig.~\ref{rms_simul}). As input parameters we used the known orbital period and typical values in CVs for the rest (1\,M$_{\rm \odot}$ as primary mass, $10^{17}$\,g\,s$^{-1}$ as mass accretion rate through the disc, inner disc radius of 10$^9$\,cm and outer disc radius estimated as 0.9 times the Roche lobe radius)\footnote{Different parameter values do not influence the behaviour we want to study/show. The principle studied in this paper is based on light curve construction, not on real flickering statistics.}.

In panel A of Fig.~\ref{rms_simul} we show a simple simulated light curve with constant mass accretion rate and corresponding rms-flux relation. This simulated light curve is equivalent to the observed quiescent phase of V1504\,Cyg shown in the upper panel of Fig.~\ref{pds_all}. The typical rms-flux relation is clear and corresponds to the observed relation.

In panel B we constructed a rising light curve by adding small light curve subsamples one by one with increasing mass accretion rate and proportionally higher numbers of flares\footnote{In our algorithm we can not change the parameters continually.}. The basic idea is based on the assumption that the mass accretion rate scales linearly with the transported angular momentum and thus the number of flickering flares. We gradually changed the mass accretion rate from 1 to $9 \times 10^{17}$\,g\,s$^{-1}$ and the number of flares from 10000 to 90000. Clearly, the resulting light curve shows the typical linear rms-flux relation. This simulation is equivalent to the rising flux of MV\,Lyr studied by \citet{scaringi2012a}.

Panel C is the same as panel B, without increasing the number of flares. The rms-flux relation is different from the two previous cases and is more similar to the observed outburst case of V1504\,Cyg shown in lower panel of Fig.~\ref{pds_all}.

In panel D we simulate the outburst light curve with typical gaps between the maxima. We made the same simulations as in panel B, but we redistributed the light curve subsamples randomly to mimic the behaviour of the observed data in Fig.~\ref{lcs}. The resulting rms-flux relation is present as expected.

The light curve in panel E is generated in the same way as in panel C, i.e. every light curve subsample is generated with the same number of flares. To get higher flux during the outbursts we added a randomly selected constant level to every small light curve segment. The resulting rms-flux relation is similar to case C, and to the observed outbursts of V1504\,Cyg.

For both simulated outburst light curves (panels D and E) we added a panel with linear fits to every individual outburst light curve segment. Not every one, but the majority of the fits have the rising trend, which agrees with the observations. The presence of decreasing linear fits in simulations or observations are due to short analysed data sets. In long enough light curves we never noticed such deviation (panel A).

\section{Discussion}
\label{discussion}

In this paper we analysed \kepler\ data of the DN V1504\,Cyg, focusing on PDS and rms-flux relation differences between quiescence and regular outbursts. Further, we investigated the evolution during the supercycle.

\subsection{PDS}

Both the quiescent and outburst PDS have a multicomponent shape with two break frequencies, with a lower and a higher frequency break, $f_{\rm L}$ and $f_{\rm H}$ respectively. It is believed that these break frequencies are the imprint of the underlying physical processes and disc structures (\citealt{scaringi2012b}, \citealt{dobrotka2015}). We found discrepant values of $f_{\rm L}$ for quiescence and outburst episodes, while $f_{\rm H}$ agrees within the errors in both PDS. Therefore, $f_{\rm L}$ should represent a structure or process which is changing during the viscous-thermal instability while $f_{\rm H}$ should arise from stable structures, unaffected by the outbursts.

Following the disc instability model (see \citealt{lasota2001} for a review) which describes the viscous-thermal instability, the mass accretion rate during the quiescence is too low causing inefficient cooling in the innermost parts of the disc. This is the reason for gas evaporation, and an optically thin geometrically thick corona forms (see e.g. \citealt{meyer1994}). On the other hand, during the outburst and hot ionized steady state of nova like systems, the mass accretion rate is strong enough, the density in the central parts ensures efficient cooling and the disc is (almost) not truncated. One would expect that the hot X-ray corona should be absent, but a sandwiched model is a possibility as proposed by \citet{scaringi2012b}, i.e. the geometrically thin disc is fully developed down to the white dwarf surface with a corona above the disc material developed from the white dwarf surface to a certain radius lower than the outer edge of the geometrically thin disc. Therefore, the inner radius of the geometrically thin disc should be different between quiescence and outburst, while the hot corona can be present during both stages. This suggests that the frequency $f_{\rm L}$ is generated in the thin disc or its inner edge, while the frequency $f_{\rm H}$ in the corona. A similar interpretation was proposed for RU\,Peg in quiescence where X-ray observations directly suggest that the low frequency is generated in the inner disc (\citealt{dobrotka2014})\footnote{We do not investigate the V1504\,Cyg PDS by simulations in this paper, because our method is not yet adapted to quiescent discs (Dobrotka, in preparation), nor to non-steady states during outbursts. Future work is planned to perform all three simulations together, i.e. flickering generated by 1) the whole quiescent disc, 2) unstable mass accretion rate from the inner quiescent disc as in RU\,Peg in \citet{dobrotka2014} and 3) full disc during outburst (with the assumption that the disc is fully developed down to the white dwarf surface during the outburst).}. Moreover, the inner disc is moving inwards during the outburst where the typical time scales are shorter (Dobrotka, in preparation, \citealt{dobrotka2012}). Therefore, a higher break frequency during outburst is plausible.

Detailed investigations of the two break frequencies in quiescence did not show any significant variability during the supercycle. We can even say, that $f_{\rm H}$ is stable with high confidence as the measurement error decreases with the investigated frequency. Therefore, the disc structure/shape should be stable during the supercycle, or hardly detectable in a PDS measurement.

The situation in outburst is slightly different. While all values of $f_{\rm L}$ agree within the errors during the supercycle, a rising trend is still possible. But the measurement error does not allow stronger conclusions. Furthermore, the frequencies $f_{\rm H}$ do not agree within the calculated 1-$\sigma$ errors, which suggests a rather unstable coronal structure. This conclusion is not surprising because we expect a corona to be present for low accretion rate during quiescence (see e.g. \citealt{meyer1994}), while a high mass accretion rate during the outburst suggests the opposite. But the corona is still possible in low density boundary regions above the disc photosphere like in the Sun, which can explain the sandwiched model. Finally, in order to strengthen our interpretation, direct X-ray detection of $f_{\rm H}$ would be needed. As described in Section~\ref{xray_try}, the X-ray count rate in available \xmm\ data is too low for meaningful conclusions.

\subsection{Rms-flux relation}

In \citet{dobrotka2015} we investigated the rms-flux relation of the nova like CV MV\,Lyr. We found that the rms-flux relation is generated by a varying number of superposed flickering flares over the simulated light curve, because the the rms and the flux increase with the number of flares. Such local increases of the number of flares can be due to a random redistribution of the flares or by the larger mass accretion rate. The former is shown in simulations shown in panel A in Fig.~\ref{rms_simul}, while the latter in panel B. The model used is not adequate for the purpose of this paper, but the principle of light curve construction is independent of the underlying physics. Panel A can be a representation of a quiescent light curve as shown in Fig.~\ref{lcs} with the corresponding rms-flux relation shown in Fig.~\ref{pds_all}. Panel B has a rising trend with clearly noticeable rising variability amplitude. This simulated light curve resembles very well the \kepler\ data of MV\,Lyr in Fig.~1 of \citet{scaringi2014} for intervals 150-450, 800-900 or 1000-1100 days. While \citet{scaringi2012a} show the rms-flux relation only for a small light curve subsample, we show the relation for the full simulated light curve. But it is valid also for small subsamples.

However, the MV\,Lyr system is a nova like CV with the disc in an ionized hot state. The same is valid for the outburst. Therefore, we expected a similar behaviour. Panel D of Fig.~\ref{rms_simul} shows an outburst light curve generated with this assumption. The discrepancy with observations is clear. In order to get the observed rms-flux relation in simulated outburst data, the light curve should be constructed in a different way. Panels C and E show the case with a constant number of flares but with added constant flux vertical offset. Therefore, the original assumption that is valid for the nova-like system MV\,Lyr (the mass accretion rate increases with the number of turbulent events) is needed to transport the enhanced angular momentum, is violated. Apparently both hot ionized states are not identical. Whether the disc is in a steady or non-steady state is essential while the activity stage does not matter. The physical reason for this different behaviour is not known to us, but at least it suggests that the quantity of flickering flares does not depend on the mass accretion rate during non steady-state outbursts while it does in steady-state discs of nova like systems. A non-turbulent steady radiation source during the outburst but not present or not dominant in the steady state of nova like CVs would resolve the puzzle.

The disc instability model as a driving mechanism of DN outbursts is described in detail in \citet{lasota2001}. Their Fig.~10 shows the surface density evolution during the outburst process. Apparently the density in the innermost parts of the disc as a dominant flickering source (see e.g. \citealt{dobrotka2010}, \citealt{dobrotka2012}) increases by two orders of magnitudes during a relatively short time. This abrupt change in flow characteristics is probably the cause for significantly different physical conditions compared to the steady state accretion of nova like systems. Therefore, the radial dependence of the mass accretion rate in the disc is probably not the only difference between outbursting and nova-like discs, but also the hydrodynamical conditions are significantly different. The turbulent characteristics of the outbursting flow is somehow saturated yielding a very similar almost equal rms value irrespective of the total flux. Therefore, the study of the enigmatic viscosity in accretion discs should be oriented in three different regimes instead of two (cold not ionized, hot ionized stages), i.e. quiescence, outburst and steady state accretion of hot ionized disc in nova like systems.

Finally the individual outburst light curves show the rms-flux relation in the majority of cases in both the simulations and observations. Some deviations from the typical rising linear relation are clear in both cases due to the short light curve duration. This rms-flux relation is generated by the random flare redistribution which is in agreement with the original assumption of underlying turbulent accretion.

\section{Summary}
\label{summary}

We analysed \kepler\ data of the dwarf nova V1504\,Cyg. We searched for differences between optical flickering in quiescence and outburst. The results can be summarized as follows:

(i) Both power density spectra show a multicomponent behaviour with two break frequencies.

(ii) The low break frequency is higher in outburst data while the higher break frequency agrees within the 1-$\sigma$ errors.

(iii) The break frequencies are stable during supercycle except for the high frequency in outburst light curves which show no significant trend.

(iv) We conclude that the low frequency break originates from regions affected by the outbursts while the stability of the higher break frequency indicates an origin that is not affected by outbursts.

(v) We propose that the low frequency break can be associated with the accretion disc or its inner edge, while the high frequency break can be generated within the inner hot geometrically thick corona. We are planning to study this model by detailed simulations of the power density spectra in a future work.

(vi) The rms-flux relation is confirmed in quiescence but not in outburst when analysing all data as a single data set. Individual outburst light curves show the typical linear rms-flux relation.

(vii) The same as previous point is valid for individual supercycle subsamples and no significant monotone evolution trend of the flux gradient is found.

(viii) We simulated both the observed quiescent and outburst rms-flux relations to test various scenarios. In the quiescent case a random redistribution of flickering flares in the simulated light curve is generating the rms-flux relation while for outbursts the basic idea was based on the assumption that the number of superposed flickering flares is proportional to the mass accretion rate. This we found valid for the hot disc in the nova-like system MV\,Lyr but not for hot discs in outburst. For the latter, the number of flickering flares must be kept unchanged, but a constant flux to the light curve must be added.

(viii) While nova-like steady state discs and non-steady state discs in dwarf novae in outburst are both in a hot ionized state, the radiation generation mechanism is not identical. The previous point suggests that flickering in both cases is generated by the turbulent mass accretion in a disc, but during the outbursts another additional significant non-turbulent source of radiation must be present.

\section*{Acknowledgement}

AD is grateful to the Slovak grant VEGA 1/0511/13. We also thank J.H.M.M. Schmitt for giving access to the currently proprietary \xmm~data.

\bibliographystyle{mn2e}
\bibliography{mybib}

\label{lastpage}

\end{document}